\documentclass[pre,a4paper,amsmath,amssymb,amsfonts,twocolumn,showpacs]{revtex4}

\usepackage{epic,eepic}
\usepackage{graphicx}
\usepackage{subfigure}

\newcommand{\br}{\ensuremath{\mathbf{r}}}
\newcommand{\bs}{\ensuremath{\mathbf{s}}}
\newcommand{\bt}{\ensuremath{\mathbf{t}}}

\newcommand{\bk}{\ensuremath{\mathbf{k}}}
\newcommand{\be}{\ensuremath{\mathbf{e}}}
\newcommand{\bel}{\ensuremath{\mathbf{e}_l}}
\newcommand{\bb}{\ensuremath{\mathbf{b}}}
\newcommand{\nbk}{\ensuremath{n^{(\mathbf{k})}}}

\newcommand{\bFex}{\ensuremath{\beta \mathcal{F}_{{\rm ex}}}}
\newcommand{\Z}{\ensuremath{\mathbb{Z}}}
\newcommand{\ZZ}{\ensuremath{\mathbb{Z}^2}}
\newcommand{\ZZZ}{\ensuremath{\mathbb{Z}^3}}
\newcommand{\ZZZZ}{\ensuremath{\mathbb{Z}^4}}

\newcommand{\dnv}{
\setlength{\unitlength}{0.1em}
\begin{picture}(8,10)(-4,3)
\path(0,0)(0,10)
\put(0,0){\whiten\circle{4}}
\put(0,10){\whiten\circle{4}}
\end{picture}}

\newcommand{\dnh}{
\setlength{\unitlength}{0.1em}
\begin{picture}(18,6)(0,-2)
\path(4,0)(14,0)
\put(4,0){\whiten\circle{4}}
\put(14,0){\whiten\circle{4}}
\end{picture}}

\newcommand{\dnp}{
\setlength{\unitlength}{0.1em}
\begin{picture}(8,6)(0,-2)
\put(4,0){\whiten\circle{4}}
\end{picture}}

\newcommand{\dnx}{
\setlength{\unitlength}{0.1em}
\begin{picture}(12.5,6)(0,0)
\path(4,0)(9.3,5.3)
\put(4,0){\whiten\circle{4}}
\put(9.3,5.3){\whiten\circle{4}}
\end{picture}}

\newcommand{\dntp}{
\setlength{\unitlength}{0.1em}
\begin{picture}(18,10)(-4,2)
\path(0,0)(10,0)(5,8.7)(0,0)
\put(0,0){\whiten\circle{4}}
\put(10,0){\whiten\circle{4}}
\put(5,8.7){\whiten\circle{4}}
\end{picture}}

\newcommand{\dntm}{
\setlength{\unitlength}{0.1em}
\begin{picture}(18,10)(-4,2)
\path(0,8.7)(10,8.7)(5,0)(0,8.7)
\put(10,8.7){\whiten\circle{4}}
\put(0,8.7){\whiten\circle{4}}
\put(5,0){\whiten\circle{4}}
\end{picture}}

\newcommand{\dntr}{
\setlength{\unitlength}{0.1em}
\begin{picture}(14,10)(-4,2)
\path(0,0)(5,8.7)
\put(0,0){\whiten\circle{4}}
\put(5,8.7){\whiten\circle{4}}
\end{picture}}

\newcommand{\dntl}{
\setlength{\unitlength}{0.1em}
\begin{picture}(14,10)(0,2)
\path(10,0)(5,8.7)
\put(10,0){\whiten\circle{4}}
\put(5,8.7){\whiten\circle{4}}
\end{picture}}

\begin{document}
\title{Density functional theory for nearest-neighbor exclusion
lattice gasses in two and three dimensions}
\date{\today}
\author{Luis Lafuente}
\email{llafuent@math.uc3m.es}

\author{Jos\'e A.\ Cuesta}
\email{cuesta@math.uc3m.es}

\affiliation{Grupo Interdisciplinar de Sistemas Complejos (GISC),
Departamento de Matem\'aticas,%Escuela Polit\'ecnica Superior,%
Universidad Carlos III de Madrid, Avda.\ de la Universidad 30,
28911--Legan\'es, Madrid, Spain}

\begin{abstract}
To speak about fundamental measure theory obliges to mention
dimensional crossover. This feature, inherent to the systems
themselves, was incorporated in the theory almost from the beginning.
Although at first it was thought to be a consistency check for
the theory, it rapidly became its fundamental pillar, thus becoming
the only density functional theory which possesses such a property. It is
straightforward that dimensional crossover connects, for instance, the
parallel hard cube system (three-dimensional) with that of squares 
(two-dimensional) and rods (one-dimensional). We show here that there
are many more connections which can be established in this way.
Through them we deduce from the 
functional for parallel hard (hyper)cubes in the
simple (hyper)cubic lattice the corresponding functionals for
the nearest-neighbor exclusion lattice gases in the square, triangular,
simple cubic, face-centered cubic, and body-centered cubic lattices.
As an application, the bulk phase diagram for all these systems is obtained.
\end{abstract}


\maketitle

\section{Introduction}

Rosenfeld's fundamental measure (FM) theory \cite{rosenfeld:1989}
is peculiar among the weighted density approximations (WDAs), and
it is so for many reasons. To begin with, classical WDAs
are constructed upon the knowledge of the thermodynamics and
structure of the uniform fluid, while FM theory is constructed on
a geometrical basis. Originally it needed scaled-particle
theory to produce a functional, but in its latest formulations
this is not a requirement anymore \cite{tarazona:1997,tarazona:2000},
and geometry stands as its unique ingredient. Another important
difference is that, while the extension to mixtures of classical
WDAs is far from being straightforward, the natural formulation
of FM theory is for a mixture (although it has recently been shown that
a FM theory for hard spheres does not accommodate a description for
mixtures as well as it was previously thought \cite{cuesta:2002}).
One further remarkable feature is that FM theory performs best where
classical theories are poorest: in the high density region.
It has been shown, for instance, that the description FM theory provides
for a solid is extraordinarily accurate in all its details
\cite{rosenfeld:1998,tarazona:2000}. This is probably the reason
why the belief has spread that FM theory is the best density
functional theory for the system of hard spheres. But there is
no free lunch. Such a peculiar structure makes the theory
extremely rigid, so much that it is very difficult (sometimes
impossible) to improve a particular detail without spoiling
another. This shows up very clearly if one tries to improve
the equation of state for the liquid phase. FM theory yields
the scaled particle equation of state. The large difference in
accuracy between the liquid and the solid gives rise to a 
not very good prediction of freezing \cite{tarazona:2000}.
If one tries to replace the equation of state by, e.g.\
Carnahan-Starling, the internal structure of the theory
squeaks \cite{tarazona:2002} and looses some of its nice
features (although for some purposes
the defects may be mostly irrelevant \cite{roth:2002}).

But by far the most genuine property of FM theory is
dimensional crossover, something that this theory shares with the
exact functionals and with no other known density functional
theory. In its first formulation this property was introduced as the
ability of a modification of Rosenfeld's original functional 
for hard spheres to recover the
exact functional for `zero-dimensional' cavities (cavities
which cannot hold more
than one sphere) \cite{rosenfeld:1996}, but it was immediately
extended to describe the property of the exact $d$-dimensional
functional to reproduce the exact $(d-1)$-dimensional one
when evaluated at a density profile which is delta-like on
a hyperplane \cite{rosenfeld:1997}. Needless to say that
functionals having dimensional crossover are particularly
suitable for studying fluids under strong confinement.

The first modified FM functional
for hard spheres produced accurate functionals for $d=2$ and
$d=1$, apart from yielding the exact one for $d=0$ (cavities)
\cite{rosenfeld:1997}. When applied to the system of parallel
hard cubes, whose FM functional can be
obtained for arbitrary dimension (being exact in $d=1$)
\cite{cuesta:1997a}, it was shown that dimensional crossover
consistently transforms the $d$-dimensional functional
into the $(d-1)$-dimensional one, down to $d=0$. The
acknowledgment that this property lies at the heart of
the formal theory suggested the last step in this direction:
transforming this property into the constructive principle
of FM theory \cite{tarazona:1997,tarazona:2000}. Under this new
formulation FM theory has been generalized to systems with
soft interaction potentials \cite{schmidt:1999}, anisotropic 
hard-particle models \cite{schmidt:2001,brader:2002},
nonadditive mixtures \cite{schmidt:2002a}, lattice
gases \cite{lafuente:2002a,lafuente:2002b,lafuente:2003},
and even fluids in porous media \cite{schmidt:2002b,schmidt:2003}.

So we see that dimensional crossover was first looked at as a
very stringent constraint on density functionals and later as
a way of rising from $d=0$ and $d=1$ to $d>1$ in the construction
of FM functionals. But dimensional crossover has another use
which has hardly been exploited: one can get new systems
out of known ones. The first (and to our knowledge the only)
example of such a use was already provided by Rosenfeld et al. 
\cite{rosenfeld:1997}. By forcing hard spheres to have
their centers of mass on one out of two parallel planes separated
a distance shorter than a sphere diameter, one obtains a
binary mixture of nonadditive hard spheres with negative
nonadditivity ($2\sigma_{12}<\sigma_{11}+\sigma_{22}$, with
$\sigma_{ij}$ the center-to-center contact distance between
spheres of type $i$ and $j$). The amount of nonadditivity
depends on the distance between planes.

Physically, dimensional crossover amounts to applying an
infinite-strength external potential all over a $(d+1)$-dimensional
lattice, except in a certain $d$-dimensional
set of sites which defines the effective system.
But this is not the only way to construct functionals for
new systems out of known ones.
One can also apply an appropriate external potential at selected
sites and modify the interaction accordingly, thus obtaining
a new system without reducing dimension.
This trick has already been applied 
to obtain the exact functional for a nonadditive mixture
of hard rods in a one-dimensional lattice from that of the
additive mixture \cite{lafuente:2002b} \footnote{A similar 
trick can be employed to recover the exact functional for the 
one-dimensional system of hard rods in a lattice 
(see Ref.~\cite{lafuente:2002b}) from the exact functional
of its continuum counterpart \cite{percus:1976}, by inserting
in the latter a density profile formed by a chain of delta
spikes (P.\ Tarazona, private communication).}. 

One can consider all systems which are related through the
kind of transformations we have just
described above. Then, because of the internal consistency 
FM theory has, the functional for a given model ``contains''
the functional for any other model to which it is related.
Our goal in this paper is
to show how this works for a set of well-known lattice
gases. The methods we will use are very general, so
their application to other families of models should not be
difficult. This is interesting if we
take into account the importance that lattice gases are
getting in the study of certain inhomogeneous problems, such as
the behavior of fluids in porous media
\cite{rosinberg:2001,rosinberg:2002,schmidt:2003}.

%Our goal in
%this paper is to exploit these devices in order to obtain
%FM functionals for new lattice gasses from those recently 
%derived for hard cubes in $d$-dimensional
%simple cubic lattices \cite{lafuente:2002b,lafuente:2003}.
%At first sight this may
%seem rather academic; however, lattice gases have been
%recently shown to have very promising applications, for
%instance, to the study of fluids in porous media 
%\cite{rosinberg:2001,rosinberg:2002,schmidt:2003}. Having
%functionals for a variety
%of different lattice gasses paves the way to
%obtaining FM functionals for more complicated models,
%like the Ising lattice gas (which includes exclusion and
%attraction), the model employed, in mean field approximation,
%in Refs.~\cite{rosinberg:2001,rosinberg:2002}.

The paper is organized as follows. Section~\ref{sec:theory} describes
the general procedure to obtain the excess free-energy
functional for nearest-neighbor exclusion lattice
gases in different lattices, either starting from a higher dimensional
functional for cubes and using dimensional reduction to a plane
or a hyperplane (this procedure is subsequently applied to 
the square, triangular, simple cubic and face-centered
cubic lattices), or starting from the functional of cubes
and applying an infinite-strength external potential in
the appropriate set of lattice sites without reducing the
effective dimensionality of the system (this procedure is
the one applied for the body-centered cubic lattice). The
final (simple) expressions for the functionals are 
explicitly obtained in closed form. In Section~\ref{sec:thermo}, these functionals
are applied to obtain the bulk phase diagram for all
the systems considered. There, FM theory results
are compared with those from other classical theories, showing
that the former is at least at the same level of accuracy than
the latter. Finally, we conclude in Section~\ref{sec:conc}.

\section{Theory}
\label{sec:theory}

In this section, we will derive the FM functional for
lattice gases with nearest-neighbor exclusion in five
different lattices: square (hard-square lattice gas), triangular
(hard-hexagon lattice gas), simple cubic (sc), faced-centered cubic (fcc)
and body-centered cubic (bcc). All these systems have been already considered 
in the literature as simple models for the hard-sphere system
\cite{runnels:1972}. As explained in the Introduction, the derivation
of the four first will make use of the dimensional
crossover property of FM functionals, so all the four will be obtained
from the known functional for $(d+1)$-dimensional parallel hard
cubes in a simple (hyper)cubic lattice \cite{lafuente:2002b,
lafuente:2003} ($d$ being the dimensionality of the final systems).
For the last one, we will start from the functional for the three-dimensional
parallel hard cubes in the simple cubic lattice, and we will apply
an infinite-strength external potential to the appropriate set of lattice sites
of the original lattice, so that
the effective lattice becomes a bcc one.
The fact that all the models considered exclude
only nearest neighbors forces the edge length of the cubes
to be $\sigma=2$ lattice spacings.

Of the different ways in which the FM functional for this
particular model of (hyper)cubes can be written \cite{lafuente:2002b,
lafuente:2003}, the simplest expression for the excess free
energy is probably
\begin{equation}\label{functional}
\bFex[\rho]=\sum_{\bs \in \Z^d}\sum_{\bk \in \{0,1\}^d} (-1)^{d-k}
\Phi_0\left(\nbk(\bs)\right),
\end{equation}
where $\nbk(\bs)$ are the weighted densities
\begin{equation}
\label{wddd}
\nbk(\bs)=\sum_{\br\in{\cal B}(\bk)}\rho(\bs+\br)
\end{equation}
labeled by the vector index $\bk=(k_1,\dots,k_d)$, $k=\sum_{l=1}^d k_l$, 
${\cal B}(\bk)$ denotes the set 
\begin{equation}\label{Bk}
{\cal B}(\bk)=\Big\{\br\in\{0,1\}^d:0\le r_i\le k_i,\ i=1,\dots,d\Big\}
\end{equation}
and $\Phi_0(\eta)=\eta+(1-\eta) \ln (1-\eta)$
is the excess free energy for a zero-dimensional cavity with mean
occupancy $0\leq\eta\leq1$ ($\beta$ is the reciprocal temperature in 
Boltzmann's constant units).

\subsection{Square lattice}
\label{sec:hs}

%This system has been studied by many authors
%before, being remarkable the treatment by using series expansions theories.
%With this technique a lot of information has been gained about the critical
%properties of the system. His phase behavior consists in
%a second order transition to an ordered state (one of the two sublattices
%is preferentially occupied) at $\rho_{{\rm crit}}=0.0000$ and
%$z_{{\rm crit}}=0.0000$.

\begin{figure}
\includegraphics[width=60mm,clip=]{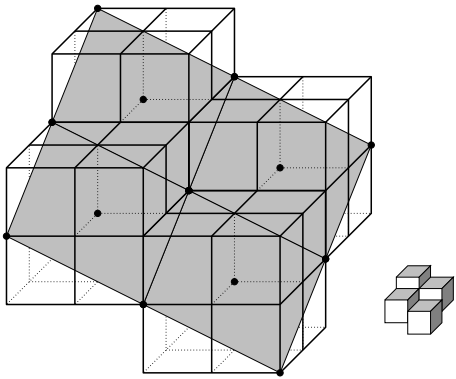}
\caption{\label{fig:rhombuses}Hard cubes with edge length $\sigma=2$
in a simple cubic lattice. Their centers of mass are constraint
to lie on the plane $s_1+s_2+2s_3=0$. Filled circles mark the
sites excluded by the cubes which lie on this plane. Notice
that on the plane, cubes behave as the hard-square lattice
gas with nearest-neighbor exclusion. The diagram at the lower
right corner helps to visualize the relative position of the
cubes in the figure.}
\end{figure}

The kind of dimensional reduction we have to perform in order to 
obtain the hard-square lattice gas out of cubes in a cubic lattice
is illustrated in Fig.~\ref{fig:rhombuses}. It amounts to forcing
the centers of mass of the cubes to lie in the plane
\begin{equation}
\label{Phs}
\mathcal{P}_{{\rm sq}}=\{(s_1,s_2,s_3)\in \ZZZ : s_1+s_2+2 s_3=0\}.
\end{equation}
(Notice that we might have chosen several equivalent planes,
given the symmetry of the system.) Figure~\ref{fig:rhombuses}
shows that the effective underlying lattice is a square lattice
defined, for instance, by the pair of orthogonal vectors
$\{\be_1=\bb_2-\bb_1,\be_2=\bb_3-\bb_1-\bb_2\}$
of the plane $\mathcal{P}_{{\rm sq}}$, $\{\bb_1,\bb_2,\bb_3\}$ being
the canonical vector basis in $\ZZZ$.
Furthermore, the effective interaction potential between the cubes
within this lattice is nearest-neighbor exclusion.

In terms of the one-particle distribution function, the dimensional
reduction can be imposed by setting
\begin{equation}
\label{dphs}
\rho(\bs)=\rho(s_2+s_3,s_3) \delta(s_1+s_2+2 s_3),
\end{equation}
where $\delta(0)=1$ and $\delta(x\ne 0)=0$ (a Kronecker symbol). The
dependence of the two-dimensional density profile reflects the choice
of basis vectors.

To obtain the FM functional for the hard-square lattice gas it only
remains to introduce the density profile (\ref{dphs}) in the functional
(\ref{functional}) through the weighted densities (\ref{wddd}) and
to eliminate the unnecessary degrees of freedom. In what follows, we
will carry out this task in detail.

The (signless) contribution of one weighted density, say $\nbk(\bs)$,
to the excess free-energy functional will be
\begin{equation*}
\sum_{\bs\in\ZZZ}\Phi_0\left(\sum_{\br\in{\cal B}(\bk)}
\rho(s_2+r_2+r_3,s_3+r_3)\delta(s_1+u)\right),
\end{equation*}
where $u\equiv r_1+r_2+2r_3$ and we have made use of the translational
invariance in $\bs$. A better way of expressing this is to
split the sum in $\bf r$ according to the values of $u$, as
\begin{equation*}
\sum_{\bs\in\ZZZ}\Phi_0\left(\sum_{u=0}^{4}\delta(s_1+u)
\sum_{(\br|u,\bk)} \rho(s_2+r_2+r_3,s_3+r_3)\right),
\end{equation*}
where $(\br|u,\bk)$ denotes those vectors $\br\in{\cal B}(\bk)$
which verify $r_1+r_2+2r_3=u$.
Now, one and only one of the delta functions within
the sum in $\br$ is 1, the others are 0, so the above expression
admits the following rewriting
\begin{equation*}
\sum_{\bs\in\ZZZ}\sum_{u=0}^{4}\delta(s_1+u)\Phi_0\left(
\sum_{(\br|u,\bk)} \rho(s_2+r_2+r_3,s_3+r_3)\right),
\end{equation*}
The sum over $s_1$ is now trivial, so denoting $\bt\equiv(s_2,s_3)$,
the expression simplifies to
\begin{equation*}
\sum_{\bt\in\ZZ} \sum_{u=0}^{4} \Phi_0\left(
\sum_{(\br|u,\bk)} \rho(s_2+r_2+r_3,s_3+r_3)\right).
\end{equation*} 
The last step concerns the identification of the argument of $\Phi_0$
(the new measures for the effective system). Using the translational
invariant in $\bt$ the above expression can
always be written
\begin{equation}
\sum_{\bt\in\ZZ} \sum_{u=0}^{4} \Phi_0\left(
\sum_{l=0}^2 a_l(\bk)\rho(\bt+{\bf e}_l)\right),
\label{wdcontribhs}
\end{equation}
where ${\bf e}_0=(0,0)$, ${\bf e}_1=(1,0)$
and ${\bf e}_2=(0,1)$, and the coefficients $a_l(\bk)$ are
listed in Table~\ref{table:hs} (notice that $\Phi_0(0)=0$, so
for some $u$ and $\bk$ there will be no contribution).

\begin{table}
\begin{ruledtabular}
\begin{tabular}{cccc}
$u$ & $a_0(\bk)$ & $a_1(\bk)$ & $a_2(\bk)$ \\
\hline
0 & 1& 0& 0 \\
1 & $k_1$& $k_2$& 0 \\
2 & $k_1 k_2$& 0& $k_3$ \\
3 & $k_1 k_3$& $k_2 k_3$& 0\\
4 & $k_1 k_2 k_3$& 0& 0 \\
\end{tabular}
\end{ruledtabular}
\caption{\label{table:hs}Coefficients of the linear combinations
$\sum_la_l(\bk)\rho(\bt+{\bf e}_l)$ appearing in (\ref{wdcontribhs}),
which define the weighted densities for the hard-square lattice gas
model.}
\end{table}

It is now time to reinsert (\ref{wdcontribhs}) back into
(\ref{functional}), group terms, and write the final form for
the functional,
\begin{equation}\begin{split}
\label{functional:hs}
\bFex^{{\rm sq}}[\rho] =&\,\sum_{\bs\in\ZZ}\Big[\Phi_0\big(\rho(\bs)+
\rho(\bs+{\bf e}_1)\big) \\
&+\Phi_0\big(\rho(\bs)+\rho(\bs+{\bf e}_2)\big)
-3\Phi_0\big(\rho(\bs)\big)\Big].
\end{split}\end{equation}
The above expression can be more easily visualized by using the
diagrammatic notation introduced in \cite{lafuente:2003},
\begin{equation}
\bFex^{{\rm sq}}[\rho]=\sum_{\bs\in\ZZ}
\left[\Phi_0\left(\dnh\right)+\Phi_0\left(\dnv\right)
-3\Phi_0\left(\dnp\right)\right].
\end{equation}

\subsection{Triangular lattice}

%The hard-hexagon model is one of the few nontrivial examples
%in two dimensions which can be exactly solved in the absence of
%external potential. The solution shows a continuous phase transition at
%$\rho^{{\rm exact}}_{{\rm crit}}=0.0000$,
%$\beta p^{{rm exact}}_{{\rm crit}}=0.0000$
%and $z^{{\rm exact}}_{{\rm crit}}=0.0000$, where one of the three
%possible sublattices is preferentially occupied. For more details
%about this exact solution we refer the reader to
%Ref.~\cite{baxter:1982}.
%
%Although the exact solution is known with no external potential,
%there are no results of this kind in the presence of arbitrary external
%fields. In this subsection, we will obtain an approximate density
%functional for this system in the same way we have done for the
%hard-square lattice gas. The benefits of having
%a functional, even an approximate one, are that this would
%open the possibility to study the hard-hexagon lattice fluid in any
%nonuniform situation,
%such as the confinement between walls, inside a porous media, etc.

\begin{figure}
\includegraphics[width=60mm,clip=]{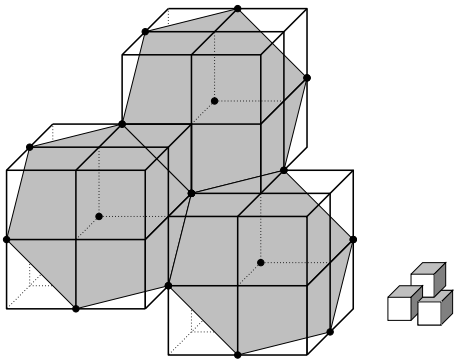}
\caption{\label{fig:hexagons}Hard cubes with edge length $\sigma=2$
in a simple cubic lattice. Their centers of mass are constraint
to lie on the plane $s_1+s_2+s_3=0$. Filled circles mark the
sites excluded by the cubes which lie on this plane. Notice
that on the plane, cubes behave as the hard-hexagon lattice
gas with nearest-neighbor exclusion. The diagram at the lower
right corner helps to visualize the relative position of the
cubes in the figure.}
\end{figure}

As in the previous subsection, we will start with the FM functional for the
parallel hard cubes with $\sigma=2$ in the simple cubic lattice.
As the procedure we will follow is the same of the one used before, we will
omit most details. Again, we will restrict the position of the centers
of mass of the cubes to lie, in this case, in the plane
\begin{equation}
\mathcal{P}_{{\rm tr}}=\{(s_1,s_2,s_3)\in\ZZZ : s_1+s_2+s_3=0\}.
\end{equation}
A sketch of the effect of this confinement is shown in 
Fig.~\ref{fig:hexagons}. It can be appreciated that the resulting
effective lattice is a triangular lattice, and that
the interaction becomes again nearest-neighbor exclusion.
The effective system thus corresponds to the hard-hexagon lattice gas.
 
The one-particle distribution function for this confined system
can be written
\begin{equation}
\rho(s_1,s_2,s_3)=\rho(s_2,s_3)\delta(s_1+s_2+s_3),
\end{equation}
where the arguments of the two-dimensional density profile
correspond to the choice $\{\be_1=\bb_2-\bb_1,
\be_2=\bb_3-\bb_1\}$ for a vector basis.

\begin{table}
\begin{ruledtabular}
\begin{tabular}{ccccc}
$u$ & $a_0(\bk)$ & $a_1(\bk)$ & $a_2(\bk)$ & $a_3(\bk)$ \\
\hline
0 & 1& 0& 0 & 0 \\
1 & $k_1$& $k_2$& $k_3$ & 0 \\
2 & 0& $k_1 k_2$& $k_1 k_3$& $k_2 k_3$ \\
3 & $k_1 k_2 k_3$& 0& 0& 0\\
\end{tabular}
\end{ruledtabular}
\caption{\label{table:hh}
The same as Table~\ref{table:hs} for the linear combinations
in (\ref{wdcontribhh}) defining
the weighted densities for the hard-hexagon lattice gas model.}
\end{table}

After eliminating the unnecessary degrees of freedom, the resulting
contribution of $\nbk(\bs)$ to the excess free-energy functional is
\begin{equation}
\sum_{\bt\in\ZZ} \sum_{u=0}^{3} \Phi_0\left(
\sum_{(\br|u,\bk)} \rho(t_1+r_2,t_2+r_3)\right),
\end{equation} 
using the same notation as in the previous case. Now
$u\equiv r_1+r_2+r_3$. This can again be written
\begin{equation}
\label{wdcontribhh}
\sum_{\bt\in\ZZ} \sum_{u=0}^{3} \Phi_0\left(
\sum_{l=0}^3 a_l(\bk) \rho(\bt+{\bf e}_l)\right),
\end{equation} 
where the coefficients $a_l(\bk)$ appear in Table~\ref{table:hh}
(a third vector, ${\bf e}_3={\bf e}_2-{\bf e}_1$, is involved
this time).

Inserting this in (\ref{functional}) and regrouping terms, the
excess free-energy functional for the hard-hexagon model turns
out to be
\begin{multline}
\label{functional:hh}
\bFex^{{\rm tr}}[\rho]=\sum_{\bs\in\ZZ}\left[
\Phi_0\left(n^{(+)}(\bs)\right)+\Phi_0\left(n^{(-)}(\bs)\right) \right.\\
\left.-\sum_{l=1}^{3}\Phi_0\left(n^{(l)}(\bs)\right)
+\Phi_0\left(\rho(\bs)\right)
\right],
\end{multline}
where we have defined the new set of weighted densities
\begin{equation}\begin{split}
n^{(\pm)}(\bs)&=\rho(\bs)+\rho(\bs\pm{\bf e}_1)+\rho(\bs\pm{\bf e}_2), \\
n^{(l)}(\bs)&=\rho(\bs)+\rho(\bs+\bel),\quad l=1,2,3.
\end{split}\end{equation}
In diagrammatic notation,
\begin{multline}
\bFex^{{\rm tr}}[\rho]=\sum_{\bs\in\ZZ}\left[
\Phi_0\left(\dntp\right)+\Phi_0\left(\dntm\right) \right.\\
%\left.-\sum_{l=1}^{3}\Phi_0\left(\dnh_{(l)}\right)
\left.-\Phi_0\left(\dnh\right)-\Phi_0\left(\dntr\right)
-\Phi_0\left(\dntl\right)+\Phi_0\left(\dnp\right)
\right].
\end{multline}

\subsection{Simple cubic lattice}

%This is the simplest lattice version of the classical
%three-dimensional hard sphere system. It has been studied before
%by series expansion methods (XXXcitasXXX). The most reasonable phase
%diagram obtained presents a second order phase transition from a disordered
%phase to an ordered one.

As in the two previous systems, we will write down a closed-form
free-energy density functional for this model, but now we have
to start from a four-dimensional system of $\sigma=2$ parallel
hypercubes in a simple hypercubic lattice and
constrain the centers of mass of the
hypercubes to be in the three-dimensional hyperplane
\begin{equation}
\mathcal{P}_{{\rm sc}}\equiv\{(s_1,s_2,s_3,s_4)\in\ZZZZ :
s_1+s_2+2s_3+4s_4=0\}.
\end{equation}
As this cannot be graphically sketched we should clarify why
we choose this particular hyperplane. A hypercube with $\sigma=2$
centered at the origin contains all lattice sites $\bs$ with
coordinates $s_l=0,\pm1$, $l=1,\ldots,4$. Now, the intersection of
such a hypercube with $\mathcal{P}_{{\rm sc}}$ can be split into
two sets: one with $s_4=0$ and $s_1+s_2+2s_3=0$, and the other one
with $s_4=\pm1$ and $s_1+s_2+2s_3=\mp4$. These two sets define three
parallel (three-dimensional) planes. The first one coincides with
$\mathcal{P}_{{\rm sq}}$, Eq.~(\ref{Phs}) (hence the choice
of the coefficients of these three coordinates in $\mathcal{P}_{{\rm sc}}$)
and thus includes the five points (including the origin) of the
two-dimensional square lattice of Sec.~\ref{sec:hs}. The last two
planes contain only one site each (symmetrically placed with
respect to the origin). Excluding the origin, these sites 
complete the eight vertices of the octahedron defining the `shape'
of the particle.

%Finally, we want to conclude this discussion by noting that the same
%argument can be used to deduce the appropriate projection
%from higher dimension than four, say $d$, in order to obtain an
%effective nearest-neighbor

The effective underlying lattice is expanded, e.g., by the vector basis
of the hyperplane $\{\be_1=\bb_2-\bb_1,
\be_2=\bb_3-\bb_1-\bb_2,\be_3=\bb_4-\bb_1-\bb_2-\bb_3\}$,
with $\{\bb_1,\bb_2,\bb_3,\bb_4\}$ the canonical vector basis of $\ZZZZ$. 
Thus, the one-particle distribution function for this system
can be written as
\begin{equation}
\rho(\bs)=\rho(s_2+s_3+2s_4,s_3+s_4,s_4)\delta(s_1+s_2+2s_3+4s_4),
\end{equation}
where, again, the three-dimensional density profile corresponds to
that of the effective system.

Repeating the procedure described in the previous examples
we arrive at this expression for the contribution of $\nbk(\bs)$
to the excess free-energy functional,
\begin{equation*}
\sum_{\bt\in\ZZZ} \sum_{u=0}^{8} \Phi_0\left(
\sum_{(\br|u,\bk)} \rho(t_1+v,t_2+r_3+r_4,t_3+r_4)\right),
\end{equation*}
where now $u\equiv r_1+r_2+2r_3+4r_4$ and $v$ is a shorthand
for $r_2+r_3+2r_4$. As in the previous cases, this becomes
\begin{equation}
\label{wdcontrib4dsc}
\sum_{\bt\in\ZZZ} \sum_{u=0}^{8} \Phi_0\left(\sum_{l=0}^3
a_l(\bk)\rho(\bt+{\bf e}_l)\right),
\end{equation}
with ${\bf e}_0=(0,0,0)$, ${\bf e}_1=(1,0,0)$, ${\bf e}_2=(0,1,0)$ and
${\bf e}_3=(0,0,1)$, and the coefficients $a_l(\bk)$ given in
Table~\ref{table:4dsc}.

\begin{table}
\begin{ruledtabular}
\begin{tabular}{ccccc}
$u$ & $a_0(\bk)$ & $a_1(\bk)$ & $a_2(\bk)$ & $a_3(\bk)$ \\
\hline
0 & 1& 0& 0 & 0 \\
1 & $k_1$& $k_2$& 0 & 0 \\
2 & $k_1 k_2$& 0& $k_3$& 0 \\
3 & $k_1 k_3$& $k_2 k_3$& 0& 0\\
4 & $k_1 k_2 k_3$& 0& 0& $k_4$\\
5 & $k_1 k_4$& $k_2 k_4$& 0& 0\\
6 & $k_1 k_2 k_4$& 0& $k_3 k_4$& 0\\
7 & $k_1 k_3 k_4$& $k_2 k_3 k_4$& 0& 0\\
8 & $k_1 k_2 k_3 k_4$& 0& 0& 0\\
\end{tabular}
\end{ruledtabular}
\caption{\label{table:4dsc}
The same as Table~\ref{table:hs} for the linear combinations
in (\ref{wdcontrib4dsc}) defining the weighted densities for the
nearest-neighbor exclusion lattice gas in the sc lattice.}
\end{table}

The final expression for the excess free-energy density functional of
this model is
\begin{equation}
\label{functional:sc}
\bFex^{{\rm sc}}[\rho]=\sum_{\bs\in\ZZZ}\left[
\sum_{l=1}^{3}\Phi_0\left(n^{(l)}(\bs)\right)
-5\Phi_0\left(\rho(\bs)\right)\right],
\end{equation}
where $n^{(l)}(\bs)=\rho(\bs)+\rho(\bs+\bel)$, or diagrammatically,
\begin{multline}
\bFex^{{\rm sc}}[\rho]=\sum_{\bs\in\ZZZ}\left[
\Phi_0\left(\dnv\right)+\Phi_0\left(\dnh\right) \right.\\
\left.+\Phi_0\left(\dnx\right)
-5\Phi_0\left(\dnp\right)\right].
\end{multline}

%It is worth mentioning that this functional recovers the two-dimensional
%one [Eq.~(\ref{functional:hs})] upon dimensional reduction to an
%appropriate plane. Moreover, all the three functionals derived by
%this procedure recover the exact one-dimensional functional for
%hard rods of length $\sigma=2$ upon reduction to a line. Thus we
%see that dimensional crossover manifests a high internal consistency
%within FM theory.

\subsection{Face-centered cubic lattice}

The nearest-neighbor exclusion lattice gas in the fcc lattice can
be obtained from the same four-dimensional system we have used
above, but now the centers of mass of the hypercubes are confined
to the three-dimensional hyperplane
\begin{equation}
\mathcal{P}_{{\rm fcc}}\equiv\{(s_1,s_2,s_3,s_4)\in\ZZZZ :
s_1+s_2+s_3+2s_4=0\}.
\end{equation}
A vector basis expanding the effective un\-der\-ly\-ing
three-di\-men\-sio\-nal 
fcc lattice is $\{\be_1=\bb_2-\bb_1,\be_2=\bb_2-\bb_3,
\be_3=\bb_4-\bb_1-\bb_3\}$. When a $\sigma=2$
hypercube is placed at the origin $(0,0,0,0)$ the excluded sites in
$\mathcal{P}_{\rm{fcc}}$ correspond to the set of twelve nearest
neighbors of the fcc lattice, whose coordinates in the chosen basis
are $\{\mathbf{e}_1=(1,0,0), \mathbf{e}_2=(0,1,0),
\mathbf{e}_3=(0,0,1), \mathbf{e}_4=(1,-1,0), \mathbf{e}_5=(1,0,-1),
\mathbf{e}_6=(0,1,-1) \}$ and the opposite ones.
Under this external potential the one-particle distribution function
takes the form 
\begin{equation}
\rho(\bs)=\rho(s_2+s_3+s_4,-s_3-s_4,s_4)\delta(s_1+s_2+s_3+2s_4),
\end{equation}
where the density profile in the r.h.s.\ is expressed in the
chosen basis.

\begin{table}
\begin{ruledtabular}
\begin{tabular}{ccccccc}
$u$ & $a_0(\bk)$ & $a_1(\bk)$ & $a_2(\bk)$ & $a_3(\bk)$ & $a_4(\bk)$ & $a_5(\bk)$\\
\hline
0 & 1& 0& 0 & 0 & 0 & 0 \\
1 & $k_1$& $k_2$& 0 & 0 & $k_3$ & 0 \\
2 & $k_1 k_3$& $k_2 k_3$& $k_1 k_2$& $k_4$& 0 & 0 \\
3 & $k_1 k_4$& $k_2 k_4$& 0& 0 & $k_3 k_4$& $k_1 k_2 k_3$ \\
4 & $k_1 k_3 k_4$& $k_2 k_3 k_4$& $k_1 k_2 k_4$& 0 & 0 & 0 \\
5 & $k_1 k_2 k_3 k_4$& 0& 0 & 0 & 0 & 0 \\
\end{tabular}
\end{ruledtabular}
\caption{\label{table:4dfcc}
The same as Table~\ref{table:hs} for the linear combinations
in (\ref{wdcontrib4dfcc}) defining
the weighted densities for the nearest-neighbor exclusion
lattice gas in the fcc lattice.}
\end{table}

Proceeding as in the previous cases, the contribution of the weighted
density $\nbk(\bs)$ to the excess free-energy functional reads
\begin{equation*}
\sum_{\bt\in\ZZZ} \sum_{u=0}^{5} \Phi_0\left(
\sum_{(\br|u,\bk)} \rho(t_1+v,t_2-r_3-r_4,t_3+r_4)\right),
\end{equation*}
where now $u\equiv r_1+r_2+r_3+2r_4$ and $v$ is a shorthand
for $r_2+r_3+r_4$. This is more conveniently rewritten
\begin{equation}
\label{wdcontrib4dfcc}
\sum_{\bt\in\ZZZ} \sum_{u=0}^{5} \Phi_0\left(\sum_{l=0}^5
a_l(\bk)\rho(\bt+{\bf e}_l)\right),
\end{equation}
with the coefficients $a_l(\bk)$ given in Table~\ref{table:4dfcc}.

Gathering together the contributions of all weighted densities,
as prescribed in (\ref{functional}), the excess free-energy density
functional of this model turns out to be
\begin{multline}
\label{functional:fcc}
\bFex^{{\rm fcc}}[\rho]=\sum_{\bs\in\ZZZ}\left[
\Phi_0\left(n^{(+)}(\bs)\right)+\Phi_0\left(n^{(-)}(\bs)\right) \right.\\
\left.
-\sum_{l=1}^{6}\Phi_0\left(n^{(l)}(\bs)\right)
+5\Phi_0\left(\rho(\bs)\right)\right],
\end{multline}
where
\begin{equation}\begin{split}
n^{(\pm)}(\bs)&=\rho(\bs)+\rho(\bs\pm{\bf e}_1)+\rho(\bs\pm{\bf e}_2)
+\rho(\bs\pm{\bf e}_3), \\
n^{(l)}(\bs)&=\rho(\bs)+\rho(\bs+\bel),\quad l=1,\ldots,6.
\end{split}\end{equation}

The diagrammatic notation is of little help in this case, so we omit
it.
%
%Diagrammatically,
%\begin{multline}
%\bFex^{{\rm fcc}}[\rho]=\sum_{\bs\in\ZZZ}\left[
%\Phi_0\left(\dnfccp\right)+\Phi_0\left(\dnfccm\right) \right.\\
%\left.
%-\sum_{l=1}^{6}\Phi_0\left(\dnh_{(l)}\right)
%+5\Phi_0\left(\dnp\right)\right],
%\end{multline}
%where subindex $(l)$ in the two-point diagram denotes the $l$th
%direction.

\subsection{Body-centered cubic lattice}

It is impossible to reduce a hypercubic four-dimensional lattice
to a three-dimensional bcc lattice by projecting
on a hyperplane. Therefore, the method used in the previous
cases is not suitable for this one.
But there exits a different procedure to obtain the functional for
this system. Again, we will begin with a system of parallel hard cubes,
but now of the same dimensionality of the target system. Also, the
key ingredient of the method is to  apply an infinite-strength
external potential to the appropriate set of lattice sites.

\begin{figure}
\includegraphics[width=60mm,clip=]{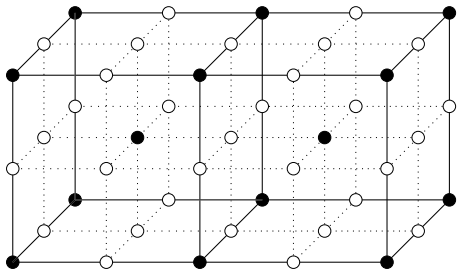}
\caption{\label{fig:bcc}Body-centered cubic lattice (black points)
obtained by applying an infinite-strength external potential at
the white points of a simple cubic lattice.
Hard cubes with $\sigma=2$ can only be placed at black points,
thus the interaction potential becomes nearest-neighbor exclusion
in the bcc lattice.}
\end{figure}

Let us consider a three-dimensional system of parallel hard cubes
with $\sigma=2$ in a simple cubic lattice. If we now restrict the
centers of mass of the cubes to the set
\begin{equation}
\mathcal{L}=\{(s_1,s_2,s_3)\in \ZZZ : \text{all $s_i$
odd or all even}\},
\end{equation}
then $\mathcal{L}$ forms a bcc lattice and the exclusion of
the cubes in the original lattice corresponds to nearest-neighbor exclusion
in the effective lattice. This transformation is sketched in Fig.~\ref{fig:bcc}.
The effect of this external potential in the density profile of the original
system amounts to
\begin{equation}
\rho(\bs)=\rho\left(\frac{s_2+s_3}{2},\frac{s_1+s_3}{2},
\frac{s_1+s_2}{2}\right)\delta_{\mathcal{L}}(\bs),
\end{equation}
where $\delta_{\mathcal{L}}(\bs)=1$ if $\bs\in\mathcal{L}$ and
$0$ otherwise, and the
coordinates in the density profile of the r.h.s.\ are referred to the
basis $\{\mathbf{e}_1=\bb_2+\bb_3-\bb_1, \mathbf{e}_2=\bb_1+\bb_3-\bb_2,
\mathbf{e}_3=\bb_1+\bb_2-\bb_3\}$, $\{\bb_1,\bb_2,\bb_3\}$ being the
canonical vector basis in $\ZZZ$.

\begin{table}
\begin{ruledtabular}
\begin{tabular}{cccccc}
$u$ & $a_0(\bk)$ & $a_1(\bk)$ & $a_2(\bk)$ & $a_3(\bk)$ & $a_4(\bk)$ \\
\hline
0 & 1& 0& 0 & 0 & $k_1 k_2 k_3$ \\
1 & $k_1$& $k_2$& $k_3$ & 0 & 0 \\
2 & $k_2$ & 0 & $k_1 k_3$& 0 & 0 \\
3 & $k_3$& 0& 0& $k_1 k_2$& 0 \\
\end{tabular}
\end{ruledtabular}
\caption{\label{table:bcc}
The same as Table~\ref{table:hs} for the linear combinations
in (\ref{wdcontrib3dbcc}) defining
the weighted densities for the nearest-neighbor exclusion
lattice gas in the bcc lattice.}
\end{table}

In order to calculate the contribution of each weighted density
$\nbk(\bs)$ we will take into account that
\begin{equation*}
\sum_{\bs\in\ZZZ}=\sum_{u=0}^{3}\sum_{\bs+\mathbf{b}_u\in\mathcal{L}},
\end{equation*}
where $\mathbf{b}_0=(0,0,0)$.
Then, after making use of the translational
invariance in $\bs$, we have for the contribution of $\nbk(\bs)$,
\begin{equation*}
\sum_{\bs\in\mathcal{L}}\sum_{u=0}^{3}\Phi_0\left(
\sum_{(\br|u,\bk)}\rho(\bs+\br-\mathbf{b}_u)\right),
\end{equation*}
where condition $(\br|u,\bk)$ selects those
$\br\in\mathcal{B}(\bk)$ such that $\br-\mathbf{b}_u\in\mathcal{L}$. 
In a more convenient way, the previous expression can be written
\begin{equation}
\label{wdcontrib3dbcc}
\sum_{\bs\in\ZZZ}\sum_{u=0}^{3} \Phi_0\left(\sum_{l=0}^{4}
a_l(\bk)\rho(\bs+\bel)\right),
\end{equation}
where $\mathbf{e}_4=\mathbf{e}_1+\mathbf{e}_2+\mathbf{e}_3$, the coefficients
$a_l(\bk)$ are given in Table~\ref{table:bcc},
and all the spatial vectors appearing
are referred to the vector basis of $\mathcal{L}$.

Taking into account the contribution of all the involved weighted densities, 
the total excess free-energy results
\begin{equation}
\label{functional:bcc}
\bFex^{{\rm bcc}}[\rho]=\sum_{\bs\in\ZZZ}\left[
\sum_{l=1}^{4}\Phi_0\left(n^{(l)}(\bs)\right)
-7\Phi_0\left(\rho(\bs)\right)\right],
\end{equation}
with $n^{(l)}(\bs)=\rho(\bs)+\rho(\bs+\bel)$. Again, the 
diagrammatic notation is of no much help in this case either.

%\begin{equation}
%\bFex^{{\rm bcc}}[\rho]=\sum_{\bs\in\ZZZ}\left[
%\sum_{l=1}^{4}\Phi_0\left(\dnh_{(l)}\right)
%-7\Phi_0\left(\dnp\right)\right].
%\end{equation}

\section{Thermodynamics}
\label{sec:thermo}

In the previous section, we have obtained the density functional
for all systems in closed form. From them it is possible to
derive all the equilibrium properties of the system in the
presence of an arbitrary external potential. In this section,
we will restrict ourselves to the bulk phase diagram. The
description of bulk behavior that lattice FM functionals
provide turns out to be equivalent to that obtained
from other classical theories. Thus in this particular application
it provides nothing new, and we include it both for completeness 
and to show the remarkable fact that FM theory subsumes many
other theories in a unified framework.
Anyhow, we want to stress that what FM theory provides are
functionals, and functionals perform best
when applied to inhomogeneous problems.

The phase transitions obtained for each model are collected
in Table~\ref{table:transitions}.

\begin{table}
\begin{ruledtabular}
\begin{tabular}{cccccc}
Lattice & Order & $\eta$ & $\beta p$ & $z$ & $-\Phi$\\
\hline
square & second & $1/2$ & 0.523 & 1.687 & 0.392 \\
triangular & first & 0.684--0.754 & 0.737 & 7.70 & --- \\
sc & second & $1/3$ & 0.305 & 0.763 & 0.350 \\
fcc & first & 0.579--0.837 & 0.461 & 5.29 & --- \\
bcc & second & $1/4$ & 0.216 & 0.490 & 0.305 \\
\end{tabular}
\end{ruledtabular}
\caption{\label{table:transitions}Transitions of nearest-neighbor
exclusion lattice gases, as predicted by FM functionals.}
\end{table}

\subsection{Square lattice}
\label{sec:hs_thermo}

As the interaction is only between nearest neighbors, we will
distinguish two sublattices in such a way that the nearest
neighbors of every site of one sublattice belongs to the
other sublattice. Let the density of each sublattice
be $\rho_1$ and $\rho_2$. Given a total density $\rho$
it holds $2\rho=\rho_1+\rho_2$. When we
consider such a density profile in (\ref{functional:hs})
the free-energy density becomes
\begin{equation}
\label{fed:sq}
\Phi^{\mathrm{sq}}=\Phi_{\mathrm{id}}+2\Phi_0(\eta)-\frac{3}{2}[\Phi_0(\rho_1)
+\Phi_0(\eta-\rho_1)],
\end{equation}
where $0\leq \eta=2 \rho \leq 1$ is the packing fraction,
$\Phi_{\mathrm{id}}=(1/2)\sum_{i}\rho_i(\ln \rho_i -1)$ is
the ideal contribution and we have written $\rho_2=\eta-\rho_1$.

The equilibrium state of the system at constant packing fraction $\eta$
is given by the global minimum of (\ref{fed:sq}) with respect to $\rho_1$.
This is a solution to the Euler-Lagrange equation
\begin{equation}
\label{el:sq}
\rho_1(1-\rho_1)^3=(\eta-\rho_1)(1-\eta+\rho_1)^3.
\end{equation}
For low values of $\eta$ this state corresponds
to the uniform or disordered one,
with $\rho_1=\rho_2=\rho$. For higher values, one expects the
system to undergo a phase transition to an ordered phase,
where one of the sublattices is preferentially
occupied. This situation is exactly the one described by the solutions of
Eq.~(\ref{el:sq}).
For $0 \leq \eta \leq \eta_{\mathrm{c}}^{\mathrm{sq}}=1/2$ the
minimum is given by $\rho_1=\rho$. Equation of state and fugacity of the
disordered phase can be easily obtained from (\ref{fed:sq}),
\begin{equation}
\beta p_{\mathrm{fluid}}^{\mathrm{sq}}=\ln \frac{(1-\eta/2)^3}{(1-\eta)^2},
\quad z_{\mathrm{fluid}}^{\mathrm{sq}}=\frac{\eta(1-\eta/2)^3}{2(1-\eta)^4}.
\end{equation}
For $1/2\leq \eta \leq 1$, we have
\begin{equation}
\rho_1^{\mathrm{eq}}=\frac{1}{2}\left[
\eta+(2-\eta)\sqrt{\frac{2\eta-1}{3-2\eta}}\right].
\end{equation}
This gives the ordered state. Equation of state and fugacity
in this phase are given by
\begin{equation}
\beta p_{\mathrm{solid}}^{\mathrm{sq}}=\beta p_{\mathrm{fluid}}^{\mathrm{sq}}
-\frac{3}{2} \ln \frac{3-2\eta}{4(1-\eta)},\quad
z_{\mathrm{solid}}^{\mathrm{sq}}=\frac{\rho_1^{\mathrm{eq}}(1-\rho_1^{\mathrm{eq}})^3}{(1-\eta)^4}.
\end{equation}

\begin{figure}
\includegraphics[width=70mm,clip=]{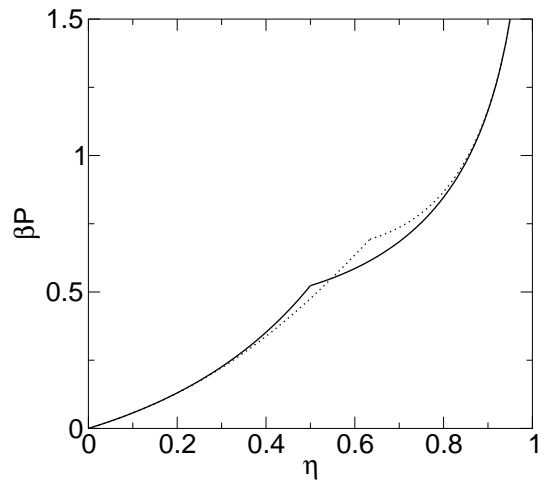}
\caption{\label{fig:eossq}Phase diagram (pressure vs.\ packing fraction)
of the hard-square lattice gas in the lattice FM approximation
(solid line). The system undergoes
a second-order phase transition from a disordered state to an
ordered one at $\eta_{\mathrm{c}}^{\mathrm{sq}}=1/2$. Dotted line: results
of the `ring' approximation obtained by Burley \cite{burley:1961}.}
\end{figure}

\begin{figure}
\includegraphics[width=70mm,clip=]{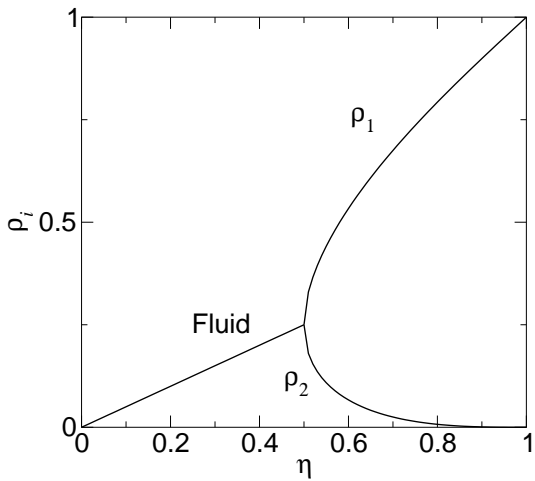}
\caption{\label{fig:cfgsq}Densities of the two sublattices, $\rho_1$
and $\rho_2$, for the hard-square lattice gas, as a function of
the packing fraction $\eta$.}
\end{figure}

The equation of state of both phases is plotted in Fig.~\ref{fig:eossq}. Also,
the occupancy of each sublattice is plotted in Fig.~\ref{fig:cfgsq}.
The transition is second order, and the critical point values 
are listed in Table~\ref{table:transitions}.
%\begin{align*}
%\eta_{\mathrm{c}}^{\mathrm{sq}}&=1/2, & \beta p_{\mathrm{c}}^{\mathrm{sq}}&=0.523, \\ 
%z_{\mathrm{c}}^{\mathrm{sq}}&=1.69 & \Phi_{\mathrm{c}}^{\mathrm{sq}}&=-0.392.
%\end{align*}
These are the same as those obtained by Burley \cite{burley:1961} and Temperley
\cite{temperley:1961} with a Bethe approximation. The next correction
to Bethe approximation (called `ring' approximation) in the scheme proposed
by Rushbrooke and Scoins \cite{rushbrooke:1955,temperley:1962a} has
also been considered by Burley \cite{burley:1961}.
The results are plotted in Fig.~\ref{fig:eossq} for comparison. The qualitative
phase behavior is the same, but the critical values improve. A further
correction (`necklace' approximation) within the same scheme has been
implemented by Temperley \cite{temperley:1962b}, who claims that
the transition becomes first order.

\subsection{Triangular lattice}

As a triangular lattice has three sublattices
such that a nearest neighbor of a site in
one sublattice belongs to the another sublattice,
we will characterize the density profile with the densities of
each sublattice: $\rho_1$, $\rho_2$ and $\rho_3$, the total
density of the system being related to them through
$3\rho=\rho_1+\rho_2+\rho_3$.

\begin{figure}
\includegraphics[width=70mm,clip=]{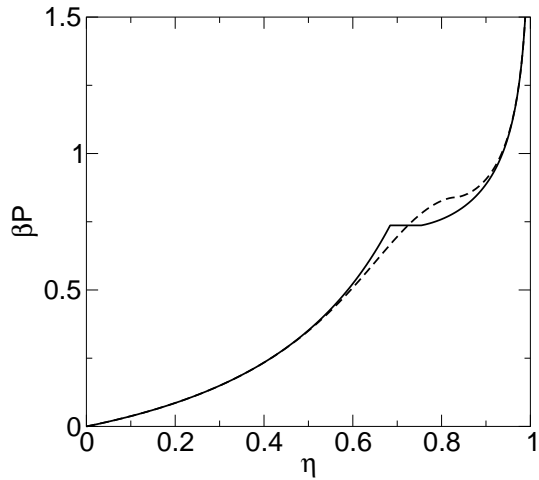}
\caption{\label{fig:eostr}The same as Fig.~\ref{fig:eossq} for the hard-hexagon
model (nearest-neighbor exclusion in a triangular lattice).
Solid line represents the equation of state from the FM approximation.
Dashed line
represents the exact solution by Baxter \cite{baxter:1982}. In our approach
the system undergoes a first-order phase transition from a disordered phase
to an ordered one. Coexisting packing fractions are
$\eta_{\mathrm{fluid}}=0.684$ and $\eta_{\mathrm{solid}}=0.754$. It is
worth mentioning the good agreement of FM approximation with the exact solution
at low and high densities, despite its failure in the critical region.}
\end{figure}

\begin{figure}
\includegraphics[width=70mm,clip=]{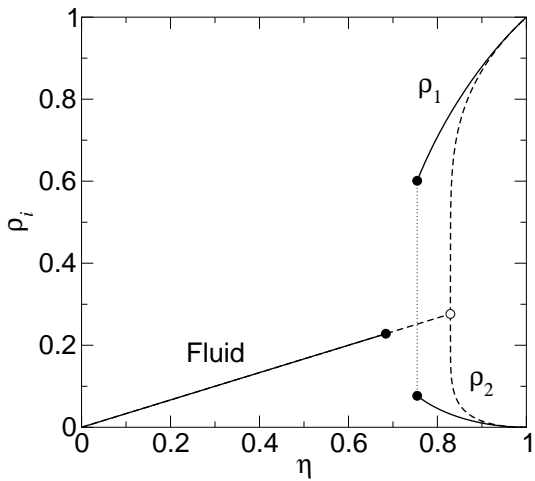}
\caption{\label{fig:cfgtr}The same as Fig.~\ref{fig:cfgsq} for the triangular
lattice. Full circles represent the configuration of coexisting phases. Dashed
lines are sublattice densities in the exact solution. Empty circle is
the exact critical point.}
\end{figure}

The free-energy density for a three sublattice configuration can be obtained
from (\ref{functional:hh})
\begin{multline}
\label{fed:tr}
\Phi^{\mathrm{tr}}=\Phi_{\mathrm{id}}+2 \Phi_0(\eta)-
\Phi_0(\rho_1+\rho_2)-\Phi_0(\rho_1+\rho_3)\\
-\Phi_0(\rho_2+\rho_3)+
\frac{1}{3}\left[\Phi_0(\rho_1)+\Phi_0(\rho_2)+\Phi_0(\rho_3)\right],
\end{multline}
where $0 \leq \eta=3\rho \leq 1$ and 
$\Phi_{\mathrm{id}}=(1/3)\sum_{i}\rho_i(\ln \rho_i -1)$.
As in the previous case, the phase diagram can be obtained by
minimizing this free-energy density at constant packing
fraction. For low values of the density the stable phase is uniform,
so $\rho_1=\rho_2=\rho_3=\rho$. In contrast, for high values of the density
one sublattice is preferentially
occupied. In this ordered phase the configuration
that results from (\ref{fed:tr})
is $\rho_1 \geq \rho_2=\rho_3$. Thus, in what follows we will consider
the latter two sublattices as equivalent. The Euler-Lagrange equation
obtained from (\ref{fed:tr}) is
\begin{equation}
\label{el:tr}
\frac{\rho_2 (1-2\rho_2)^3 (1-\eta+2\rho_2)}
{(1-2\rho_2) (\eta-2\rho_2) (1-\eta+\rho_2)^3}=1.
\end{equation}
The solutions to this equation as a function of $\eta$ indicate a
first-order transition. The coexisting packing fractions as well
as the pressure and fugacity at coexistence appear in 
Table~\ref{table:transitions}.
%\begin{equation*}
%\eta_{\mathrm{fluid}}^{\mathrm{tr}}=0.684,\quad
%\eta_{\mathrm{solid}}^{\mathrm{tr}}=0.754,
%\end{equation*}
%and
%\begin{equation*}
%\beta p^{\mathrm{tr}}=0.737,\quad z^{\mathrm{tr}}=7.70.
%\end{equation*}
The equation of state for the disordered phase is
\begin{equation}
\beta p^{\mathrm{tr}}_{\mathrm{fluid}}=\ln \frac{(1-2\eta/3)^3}
{(1-\eta/3)(1-\eta)^2},
\end{equation}
and the fugacity
\begin{equation}
z_{\mathrm{fluid}}^{\mathrm{tr}}=\frac{\eta (1-2\eta/3)^6}
{3(1-\eta/3)(1-\eta)^6}.
\end{equation}
For the ordered phase
\begin{equation}
\beta p_{\mathrm{solid}}^{\mathrm{tr}}=\frac{1}{3}
\ln \frac{(1-\eta+\rho_2)^6 (1-2 \rho_2)^3}
{(1-\eta)^6 (1-\eta+2 \rho_2) (1-\rho_2)^2},
\end{equation}
and
\begin{equation}
z_{\mathrm{solid}}^{\mathrm{tr}}=\frac{\rho_2 (1-\eta+\rho_2)^3 (1-2\rho_2)^3}
{(1-\eta)^6 (1-\rho_2)},
\end{equation}
where $\rho_2$ is given by (\ref{el:tr}).
The phase diagram is plotted in Fig.~\ref{fig:eostr}. Sublattice densities
are shown in Fig.~\ref{fig:cfgtr}.

This result coincides with that obtained by Burley \cite{burley:1965} by using
the `ring' approximation (see Sec.~\ref{sec:hs_thermo}), and was also
obtained by Temperley through a combinatorial
finite matrix method \cite{temperley:1961}. 

This system has been exactly solved by Baxter \cite{baxter:1980b,baxter:1982} (the exact
solution is shown in Figs.~\ref{fig:eostr} and \ref{fig:cfgtr}). The system exhibits a
\textsl{continuous} phase transition (the critical point in the
pressure vs.\ density diagram is a horizontal inflection point) at packing
fraction $\eta_{\mathrm{c}}^{\mathrm{exact}}=(3/10)(5-\sqrt{5})\approx0.829$
and critical pressure $\beta p_{\mathrm{c}}^{\mathrm{exact}}=(1/2)\ln[(27/250)
(25+11\sqrt{5})]\approx 0.839$. This is the first time that FM theory can be compared
with an exact result in dimension higher than one. Although the theory fails in
predicting correctly the order of the transition, the agreement in the whole range of
density except the critical region is rather accurate.

%Furthermore, a clear
%evidence shows up in this comparison: the lack of an accurate description for the
%fluid branch is responsible of the poor behavior of FM theory in the critical
%region. While the exact fluid branch attains a maximum in the critical point,
%the FM fluid branch increases monotonically. In the exact solution what
%happens is that the maximum of the fluid branch reaches the minimum of the
%solid branch, resulting the critical point. Although the FM solid branch behaves
%like the exact one (it has a minimum), this can not ever be reached by
%the fluid branch.

\subsection{Simple cubic lattice}

In the close-packed state the particles of this system
occupy one sublattice with fcc symmetry, while
the sublattice formed by nearest neighbors of that one remains empty. Thus,
we will only consider these two sublattices, with densities
$\rho_1$ and $\rho_2$, respectively. Again, the total density $\rho$
is related to these numbers by $2 \rho=\rho_1+\rho_2$.

When we insert such a density profile in the functional (\ref{functional:sc})
we get for the free-energy density
\begin{equation}
\label{fed:sc}
\Phi^{\mathrm{sc}}=\Phi_{\mathrm{id}}+\Phi_0(2\rho)
-\frac{5}{2}\left[
\Phi_0(\rho_1)+\Phi_0(2\rho-\rho_1)\right],
\end{equation}
where we have eliminated the dependency on $\rho_2$.
For a fixed value of the density, the global minimum of (\ref{fed:sc})
is a solution to the Euler-Lagrange equation
\begin{equation}
\label{el:sc}
\frac{\rho_1(1-\rho_1)^5}{(\eta-\rho_1)(1-\eta+\rho_1)^5}=1,
\end{equation}
$0\leq \eta=2\rho \leq 1$ being the packing fraction. When
$0\leq \eta \leq \eta_{\mathrm{c}}^{\mathrm{sc}}=1/3$, the minimum
is given by the uniform phase $\rho_1^{\mathrm{eq}}=\eta$. For
the equation of state and fugacity we have
\begin{equation}
\beta p_{\mathrm{fluid}}^{\mathrm{sc}}=\ln \frac{(1-\eta/2)^5}{(1-\eta)^3},\quad
z_{\mathrm{fluid}}^{sc}=\frac{\eta(1-\eta/2)^5}{2(1-\eta)}.
\end{equation}
Upon increasing the density the system experiences a second-order disorder-order
transition at $\eta_{\mathrm{c}}^{\mathrm{sc}}=1/3$.
For $1/3 \leq \eta \leq 1$, the density of the preferred sublattice can be calculated
from (\ref{el:sc}) to be
\begin{equation}
\rho_1^{\mathrm{eq}}=\frac{1}{2}\left\{
\eta+(2-\eta)\sqrt{\frac{2[5-4\eta(2-\eta)]^{1/2}-5(1-\eta)}
{5-3\eta}}\right\}.
\end{equation}
In this phase, the equation of state is given by
\begin{equation}
\beta p_{\mathrm{solid}}^{\mathrm{sc}}=
\beta p_{\mathrm{fluid}}^{\mathrm{sc}}+\frac{5}{2}
\ln \left[\frac{2(5-4\eta)-\sqrt{5-4\eta(2-\eta)}}
{5-3\eta}\right],
\end{equation}
and the fugacity by
\begin{equation}
z_{\mathrm{solid}}^{\mathrm{sc}}=\frac{\rho_1^{\mathrm{eq}}(1-\rho_1^{\mathrm{eq}})^5}
{(1-\eta)^6}.
\end{equation}

\begin{figure}
\includegraphics[width=70mm,clip=]{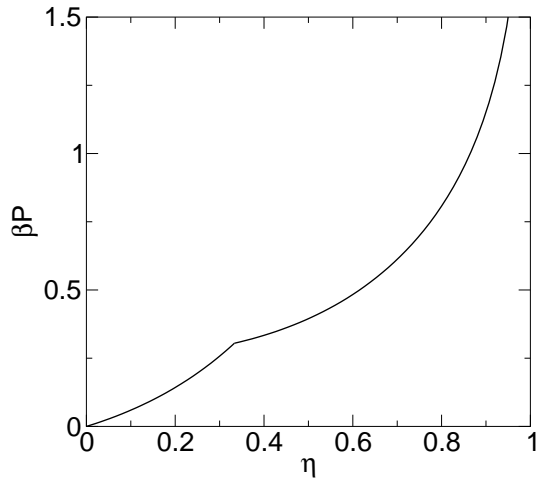}
\caption{\label{fig:eossc}The same as Fig.~\ref{fig:eossq} for the nearest-neighbor
exclusion lattice gas in the simple cubic lattice. A second-order transition is
predicted at $\eta_{\mathrm{c}}^{\mathrm{sc}}=1/3$.}
\end{figure}

\begin{figure}
\includegraphics[width=70mm,clip=]{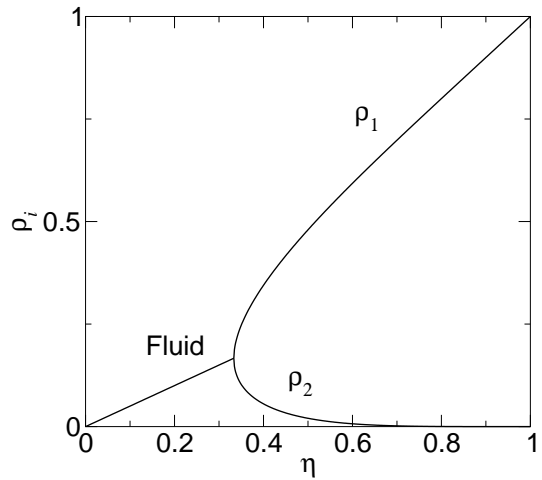}
\caption{\label{fig:cfgsc}The same as Fig.~\ref{fig:cfgsq} for the simple cubic
lattice.}
\end{figure}

The equation of state is plotted in Fig.~\ref{fig:eossc}.
Sublattice densities are shown in Fig.~\ref{fig:cfgsc}. The critical point values
are listed in Table~\ref{table:transitions}.
As for the square lattice, for this model the result from FM theory coincide
with a Bethe approximation \cite{burley:1961}.

\subsection{Face-centered cubic lattice}

An fcc lattice contains four equivalent sublattices
each with a sc symmetry. As in the previous cases
sublattice densities are denoted, $\rho_1,\rho_2,\rho_3,\rho_4$, and
$\eta=4 \rho=\rho_1+\rho_2+\rho_3+\rho_4$ holds for the total density $\rho$ or
packing fraction $\eta$.

The free-energy density for such a density profile
in the FM approximation is obtained from (\ref{functional:fcc}) as
\begin{equation}
\label{fed:fcc}
\Phi^{\mathrm{fcc}}=\Phi_{\mathrm{id}}+2 \Phi_0(\eta)
-\sum_{i<j}\Phi_0(\rho_i+\rho_j)+\frac{5}{4}\sum_{i=1}^{4}
\Phi_0(\rho_i),
\end{equation}
where $\Phi_{\mathrm{id}}=(1/4)\sum_{i} \rho_i(\ln \rho_i -1)$.
The global minimum of $\Phi_{\mathrm{fcc}}$ at fixed density
with respect to the sublattice densities yields the
thermodynamically stable phase. 
But now there is a subtle point we have to take into account:
the close-packing is degenerated.
If we consider the fcc lattice as a stacking of
square lattices each having its sites on the centers of the
squares of the previous one, then we can fill alternative
layers independently. Thus, we will consider different densities for
the four sublattices and the minimum of the free
energy will determine the structure of the most stable phase.
(Previous studies \cite{gaunt:1967} impose
the equivalence of three sublattices from the beginning,
\textit{i.e.} $\rho_1\neq\rho_2=\rho_3=\rho_4$.) 

\begin{figure}
\includegraphics[width=70mm,clip=]{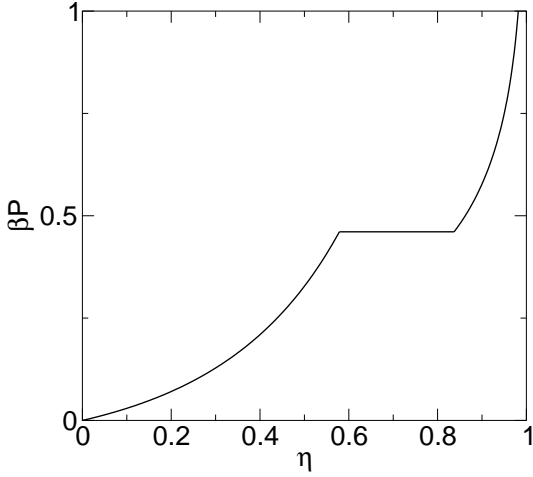}
\caption{\label{fig:eosfcc}The same as Fig.~\ref{fig:eossq} for the nearest-neighbor
exclusion in the face-centered cubic lattice. A first-order phase transition takes place
at pressure $\beta p=0.461$ with coexisting states at
$\eta_{\mathrm{fluid}}^{\mathrm{fcc}}=0.579$ and $\eta_{\mathrm{solid}}^{\mathrm{fcc}}=0.837$.}
\end{figure}

\begin{figure}
\includegraphics[width=70mm,clip=]{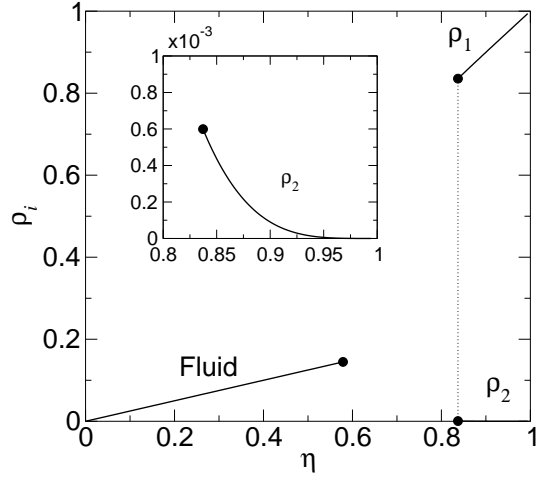}
\caption{\label{fig:cfgfcc}The same as Fig.~\ref{fig:cfgsq} for the fcc
lattice. Solid circles represent configurations of coexisting
phases. The inset shows a detail of sublattice density $\rho_2$.}
\end{figure}

Solving numerically the Euler-Lagrange equations obtained from (\ref{fed:fcc})
at fixed packing fraction, we obtain the equation of state plotted in
Fig.~\ref{fig:eosfcc}. Sublattice densities at equilibrium phases are
shown in Fig.~\ref{fig:cfgfcc}. The system undergoes a first-order phase transition
to an ordered phase with $\rho_1\neq\rho_2=\rho_3=\rho_4$. (See Table~\ref{table:transitions}
for the critical point values.)
%At the transition we have
%\begin{equation*}
%\eta_{\mathrm{fluid}}^{\mathrm{fcc}}=0.579,\quad
%\eta_{\mathrm{solid}}^{\mathrm{fcc}}=0.837,
%\end{equation*} 
%and
%\begin{equation*}
%\beta p^{\mathrm{fcc}}=0.461,\quad z^{\mathrm{fcc}}=5.29.
%\end{equation*}
For the uniform phase
\begin{equation}
\beta p_{\mathrm{fluid}}^{\mathrm{fcc}}=\ln \frac{(1-\eta/2)^6}
{(1-\eta/4)^5 (1-\eta)^2}
\end{equation}
and
\begin{equation}
z_{\mathrm{fluid}}^{\mathrm{fcc}}=\frac{\eta (1-\eta/2)^{12}}
{4 (1-\eta/4)^5 (1-\eta)^8}.
\end{equation}
For the ordered phase
\begin{equation}
\beta p_{\mathrm{solid}}^{\mathrm{fcc}}=\frac{1}{4}
\ln \frac{(1-\eta+2\rho_2)^{12} (1-2\rho_2)^{12}}
{(1-\eta+3\rho_2)^5 (1-\rho_2)^{15} (1-\eta)^8}
\end{equation}
and
\begin{equation}
z_{\mathrm{solid}}^{\mathrm{fcc}}=\frac{\rho_2 (1-\eta+2\rho_2)^4 (1-2\rho_2)^8}
{(1-\rho_2)^5 (1-\eta)^8},
\end{equation}
with $\rho_2$ the minimum of the free energy (\ref{fed:fcc}).

It is worth mentioning that a metastable phase transition from the
disordered state to a smectic ordered state (alternative square-lattice layers are
uniformly occupied, with densities $\rho_1$ and $\rho_2$) is also obtained
from the functional (\ref{fed:fcc}).

\subsection{Body-centered cubic}

At close packing state this system has one sublattice completely filled while
the nearest-neighbor sublattice is empty. Therefore, we will consider
the density profile as in the previous similar cases (square or sc lattices).

\begin{figure}
\includegraphics[width=70mm,clip=]{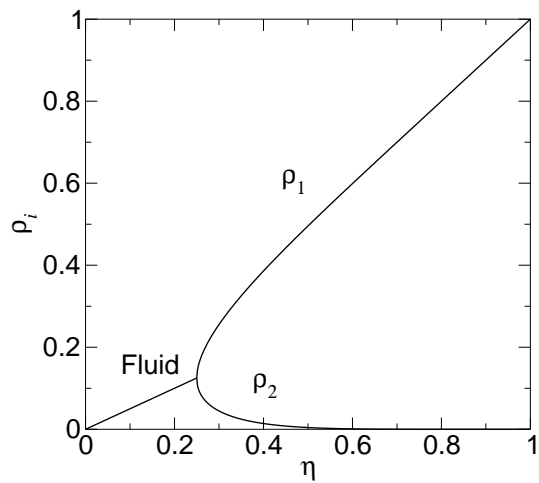}
\caption{\label{fig:cfgbcc}The same as Fig.~\ref{fig:cfgsq} for the bcc
lattice.}
\end{figure}

The free-energy density [Eq.~(\ref{functional:bcc})] becomes
\begin{equation}
\label{fed:bcc}
\Phi^{\mathrm{bcc}}=\Phi_{\mathrm{id}}+4\Phi_0(\eta)-\frac{7}{2}
\left[\Phi_0(\rho_1)+\Phi_0(\eta-\rho_1)\right].
\end{equation}
The Euler-Lagrange equation is
\begin{equation}
\label{el:bcc}
\frac{\rho_1(1-\rho_1)^7}{(\eta-\rho_1)(1-\eta+\rho_1)^7}=1.
\end{equation}
The phase behavior obtained from this equation is similar to that of the
square or the sc lattices: The system
undergoes a second-order phase transition at
$\eta_{\mathrm{c}}^{\mathrm{bcc}}=1/4$; below this density the stable phase
is a fluid, while above the particles occupy preferentially
one sublattice. The ordered solution of
(\ref{el:bcc}) can be obtained analytically, but the expression is rather
cumbersome, so we just plot it
in Fig.\ref{fig:cfgbcc}. For the fluid phase, equation of state and
fugacity are given by
\begin{equation}
\beta p_{\mathrm{fluid}}^{\mathrm{bcc}}=\ln \frac{(1-\eta/2)^7}{(1-\eta)^4},\quad 
z_{\mathrm{fluid}}^{\mathrm{bcc}}=\frac{\eta(1-\eta/2)^7}{2(1-\eta)^8}.
\end{equation}
For the ordered phase, we have
\begin{equation}
\beta p_{\mathrm{solid}}^{\mathrm{bcc}}=\frac{1}{2}\ln \frac{(1-\rho_1)^7
(1-\eta+\rho_1)^7}{(1-\eta)^4}, 
\end{equation}
and
\begin{equation}
z_{\mathrm{solid}}^{\mathrm{bcc}}=\frac{\rho_1 (1-\rho_1)^7}{(1-\eta)^8},
\end{equation}
with $\rho_1$ the ordered solution of (\ref{el:bcc}).

\begin{figure}
\includegraphics[width=70mm,clip=]{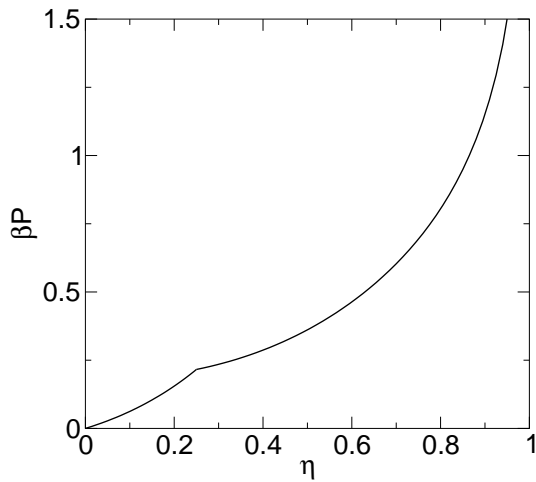}
\caption{\label{fig:eosbcc}The same as Fig.~\ref{fig:eossq}
for the nearest-neighbor exclusion lattice gas in the bcc lattice.
The system undergoes a second-order phase ordering transition at
$\eta_{\mathrm{c}}^{\mathrm{bcc}}=1/4$.}
\end{figure}

These results are again equivalent to those obtained with a Bethe approximation
\cite{burley:1961}. The equation of state is plotted in Fig.~{\ref{fig:eosbcc}}.
(See Table~\ref{table:transitions} for the critical point values)
%\begin{align*}
%\eta_{\mathrm{c}}^{\mathrm{bcc}}&=1/4, & 
%\beta p_{\mathrm{c}}^{\mathrm{bcc}}&=0.216, \\ 
%z_{\mathrm{c}}^{\mathrm{bcc}}&=0.490 &
%\Phi_{\mathrm{c}}^{\mathrm{bcc}}&=-0.305.
%\end{align*}

\section{Discussions and Conclusions}
\label{sec:conc}

Dimensional crossover is a property that connects different models. A
system of parallel hard cubes (both on and off a lattice) constrained
to lie on a plane parallel to their sides becomes
a system of parallel hard squares. A system of parallel hard squares
constrained to lie on a straight line parallel to their edges 
becomes a system of hard rods. These are obvious dimensional
crossovers, but in this paper we have introduced a few more:
parallel hard cubes constrained to lie on more general planes
produce nearest-neighbor exclusion lattice gases in the square
and triangular lattices; four-dimensional parallel hard hypercubes
constrained to lie on certain hyperplanes become nearest-neighbor
exclusion lattice gases in the sc or fcc lattices. If this
dimensional constraint is regarded as the application of
an infinite-strength external field, then more general patterns
of such an external field produce new models out of the parallel
hard cube one (the nearest-neighbor exclusion lattice gas in the
bcc lattice, for instance).

The message we want to transmit with this work is that FM
functionals for all these models have the remarkable property
of being connected with each other through the same transformations.
More than that: further dimensional crossovers of these models
are also consistently captured by FM functionals. For instance,
all functionals obtained in this paper produce the {\em exact}
functional for hard rods when reduced to one dimension; also,
the functional for the nearest-neighbor lattice gas in the
square lattice, Eq.~(\ref{functional:hs}), can be obtained
from the one in either the sc lattice, Eq.~(\ref{functional:sc}),
the fcc lattice, Eq.~(\ref{functional:fcc}), or the bcc lattice,
Eq.~(\ref{functional:bcc}), through the application of an
appropriate external potential. Similarly, the functional 
for the nearest-neighbor lattice gas in the triangular
lattice, Eq.~(\ref{functional:hh}), can be recovered from that
in the fcc lattice, Eq.~(\ref{functional:fcc}).
Such a degree of internal consistency is not shared by
any other known density functional theory, and it puts
FM theory at a different level. As a byproduct, as pointed
out in the Introduction, it warrants a good behavior of
FM functionals when dealing with highly inhomogeneous 
situations.

When applied to study the bulk phase behavior of these 
nearest-neighbor lattice gases, FM functionals produce
reasonable results, sometimes with important discrepancies
in the critical region, sometimes more accurate, and
always very accurate at low and high densities. There are,
no doubt, better methods to fit the equation of state and
obtain a better description of this phase behavior, like,
for instance, finite-size analysis \cite{runnels:1965,ree:1966,
orban:1968,guo:2002} or series expansions \cite{gaunt:1965,
gaunt:1967,orban:1968,baxter:1980a}. They are particularly
useful in predicting the critical behavior. FM approach has
an important advantage on these method, whatever the 
sacrifice in accuracy: it leads to simple, closed-form
functionals. Thus it permits to study inhomogeneous problems,
which are absolutely out of the scope of these other 
more accurate methods. A particularly interesting
example of such inhomogeneous problems can be
found in recent studies of fluids in porous media
\cite{rosinberg:2001,rosinberg:2002,schmidt:2003}, for which
lattice models seem to capture
enough physical information to describe many interesting
phenomena not yet described by other models.

An interesting observation to make from the results of
the application of FM theory to describe bulk phase behavior
is the close connection it has with other classical approaches.
For those lattices with only two sublattices (loose-packed
lattices) FM theory reduces to a Bethe approximation
(square, sc and bcc lattices);
for lattices with more than one sublattice (close-packed
lattices) it becomes equivalent to another cluster-like
approximation. In fact, when closely looked at, all these
approximation are but particular cases of 
Kikuchi's cluster variation theory \cite{kikuchi:1951}.
This theory proposes a hierarchical scheme of approximations
under the basis of describing in an exact manner
clusters of increasing size. Roughly speaking, the larger the
cluster the more accurate the results---and the more
involved the theory. The connection between FM theory and
the cluster variation method is definitely worth exploring,
because one drawback of the latter is that there is not
an {\em a priori} criterion to choose the clusters, other
than ``the larger the better''; as a matter of fact, some
clusters (depending on the model) produce an optimal result
and some others (even larger ones) spoil the accuracy, and
the reason for that is unknown. FM theory, on the contrary,
leaves no freedom to choose the clusters, but those it
prescribes seem to be optimal in the above sense. It is
very illustrative Burley's treatment of the nearest-neighbor
exclusion lattice gas in the fcc lattice \cite{burley:1961}:
the inappropriate election of the clusters [only part of
those prescribed by the FM functional (\ref{functional:fcc})]
leads to a
spurious divergence of the free-energy at a density lower
than the close-packing. This is a line of investigation
we are currently following.

\begin{acknowledgments}
This work is supported by project BFM2000-0004 from the Direcci\'on
General de Investigaci\'on (DGI) of the Spanish Ministerio de Ciencia
y Tecnolog\'{\i}a.
\end{acknowledgments}

\bibliography{../llafuent}

\end{document}